# Progress in Water-Based Metamaterial Absorber: A Review

Jingda Wen[1,2], Qiang Ren[1*], Ruiguang Peng[2], Haoyang Yao[3],

Yuchang Qing[3], Jianbo Yin[4], Qian Zhao[2*]

**Affiliations：**

1. School of Electronics and Information Engineering, Beihang University, Beijing 100191, China

2. State Key Lab of Tribology, Department of Mechanical Engineering, Tsinghua University, Beijing 100084, China

3. State Key Laboratory of Solidification Processing, Northwestern Polytechnical University, Xi'an 710072, China

4. Smart Materials Laboratory, Department of Applied Physics, Northwestern Polytechnical University, Xi'an, 710129, China.

*Author to whom correspondence should be addressed. E-mail: qiangren@buaa.edu.cn & zhaoqian@tsinghua.edu.cn

**Graphical Abstract:**

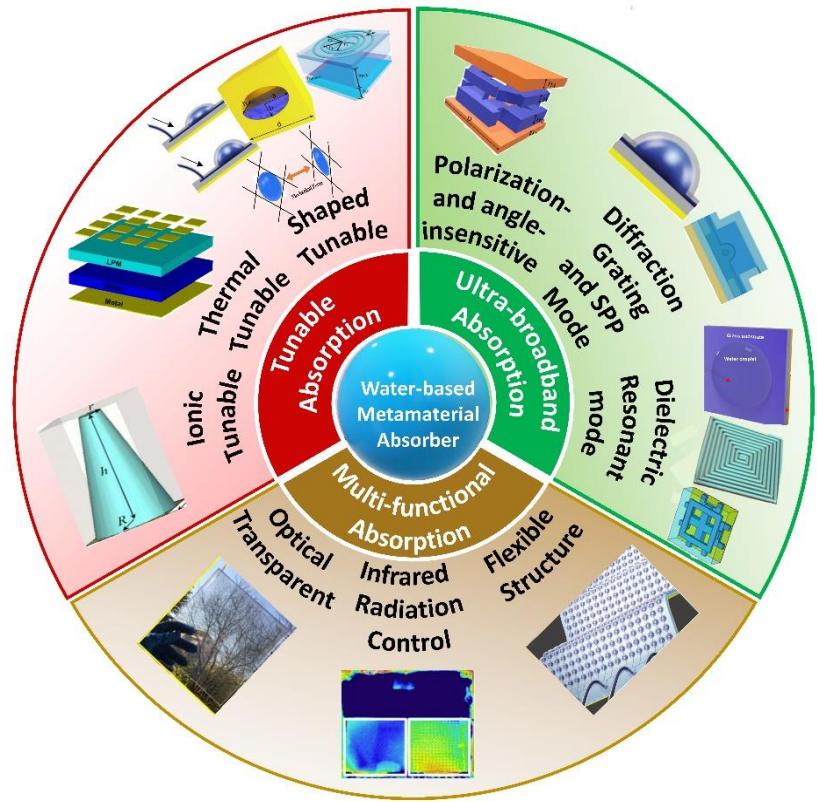

## Abstract:

Increasing attention on microwave ultra-broadband metamaterial absorbers has been paid due to their promising applications. While most microwave ultra-broadband metamaterial absorbers developed so far are based on metallic resonant structures, dispersive dielectric water-based metamaterial opens a simpler and more versatile route for the construction of polarization- and angle- insensitive ultra-broadband absorption. Here, we review the recent progresses of water-based metamaterial absorber by providing an illustration of the mechanisms to realize ultra-broadband, tunable and multi-functional absorption. We also address the further development direction and some potential novel applications.

## Introduction:

Recently, substantial progress in microwave absorption has been achieved for various novel potential applications, such as radar stealth, radiation prevention, and energy collection [1-5]. With the development of metamaterials [6-8], metamaterial absorbers can substantially dissipate microwave energy by manipulating effective permittivity and permeability parameters simultaneously based on the impedance-matching principle [9]. Due to the limitation between quality factor and resonant intensity, only high absorption at single frequency can be achieved in such designs [10]. In order to break this application limitation, coupled integration methods are proposed to construct dual-band and multi-band metamaterial absorbers by combining metamaterial absorber unit with coupled adjacent resonance peaks vertically or horizontally [11-13]. The dual-band and multi-band metamaterial absorbers have more than one ranges of absorption frequency, but the continuous broad frequency band cannot be covered in these designs [14]. Although tunable metamaterial manipulation methods, such as diodes [15], graphene [16], liquid crystal[17], electric field[18], phase change material[19] and thermal control [20], are successfully used to fabricate frequency tunable metamaterial absorbers for continuous broad frequency band absorption, the simpler ultra-broadband metamaterial absorbers are highly sought after for better novel applications.

In order to achieve ultra-broadband absorption, the strategy of employing self-similar structure to realize coupled resonance peaks is proposed in [21]. Usually, the microwave ultra-broadband metamaterial absorbers are integrated coupled self-similar metallic resonance structures with adjacent resonance peaks. High absorption can be achieved in ultra-broad frequency band by vertical and horizontal integration method. Resonance modes of absorber correspond to frequency point within absorption band, while the resonance at lower or higher frequency exists in the parts with larger or smaller dimensions, respectively. Due to the coupled resonance modes, the absorber can achieve impedance-matching condition in ultra-broad band with flat dispersion of constitutive parameters [22]. However, the metallic ultra-broadband flat dispersion structures are fabricated with complex integration and optimization process. It is important to find a simpler and more practical design method to reach the theoretical minimal thickness with high absorption in low frequency regime [23].

Most of microwave ultra-broadband metamaterial absorbers developed so far are based on metallic resonant structures, while the others are the dielectric resonant structures. Recently, the promising periodic dielectric water-droplet WMMA is proposed [24], effectively utilizing the constitutive dispersive permittivity of water. The promising water-based metamaterial with plenty of advantages, such as biocompatibility, easy availability, versatile tunability, and optical transparency, can be applied in radar stealth, energy harvesting and radiation prevention. In this paper, we focus on the potential of microwave ultra-broadband absorption based on water-based metamaterials, and analyze the mechanisms of dielectric resonance, spoof surface polarizations and diffraction gratings in WMMA. Moreover, we exhibit the recent progress in WMMAs in thermal tunability, ionic tunability, shape tunability, and multiple functions. Finally, the future development prospects and novel potential applications of microwave WMMA are discussed.

I. **Dispersive relation for better absorption performance**

The dielectric water exists as liquid from 0 to 100 °C at atmospheric pressure [25]. The charge center shift of liquid water molecules in polarized state process can be described by Debye equations. At the room temperature $T = 20°C$, the angular frequency $\omega$ and permittivity $\varepsilon = \varepsilon_{real} + j\varepsilon_{image} = \varepsilon_{real}(1 - jtan\delta)$ are shown in Eq. (1-4). The symbols $\varepsilon_\infty(\omega,T)$, $\varepsilon_s(\omega,T)$ and $\tau(\omega,T)$ are optical permittivity, static permittivity and rotational relaxation time, respectively. (More details are available in [26])

$$\varepsilon = \varepsilon_\infty(\omega,T) + \frac{\varepsilon_s(\omega,T) - \varepsilon_\infty(\omega,T)}{1 - i\omega\tau(\omega,T)} \quad (1)$$

$$\varepsilon_{real} = \varepsilon_\infty(\omega,T) + \frac{\varepsilon_s(\omega,T) - \varepsilon_\infty(\omega,T)}{1 + \omega^2\tau^2(T)} \quad (2)$$

$$\varepsilon_{image} = \frac{[\varepsilon_s(\omega,T) - \varepsilon_\infty(\omega,T)]\omega\tau}{1 + \omega^2\tau^2(T)} \quad (3)$$

$$tan\delta = \frac{[\varepsilon_s(\omega,T) - \varepsilon_\infty(\omega,T)]\omega\tau}{\varepsilon_s(\omega,T) + \varepsilon_\infty(\omega,T)\omega^2\tau^2(T)} \quad (4)$$

As shown in Fig. 1, the permittivity and loss tangent curves show that water has extremely high loss within a certain microwave frequency band, due to the energy dissipation of molecule polarization process.

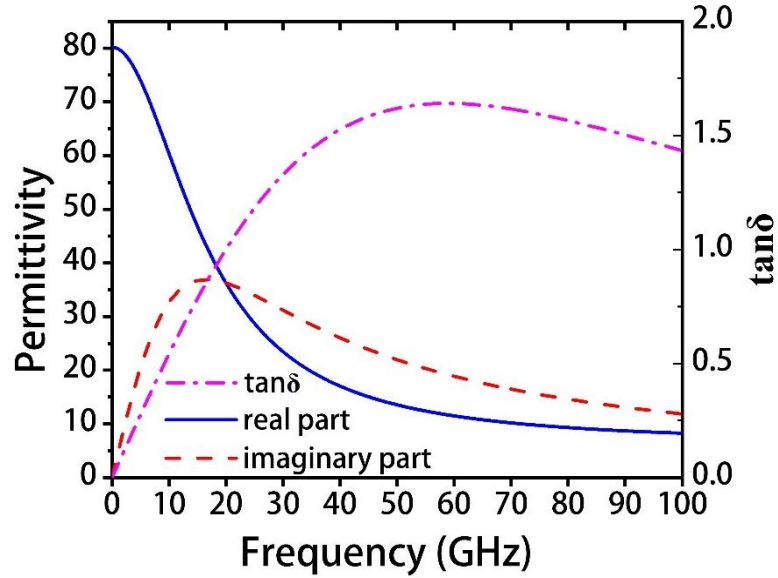

Fig. 1. The permittivity and lost tangent of water in microwave frequency regime at room temperature.

(Authors' unpublished figure)

Considering a sub-wavelength metamaterial absorber with thickness $d$ as a uniform equivalent layer, the normalized impedance $z_r$ and refractive index $n_r$ can be extracted by scattering parameters [27].

$$z_r = \pm\sqrt{\frac{(1+S_{11})^2 - S_{21}^2}{(1-S_{11})^2 - S_{21}^2}} \tag{5}$$

$$\exp(in_r kd) = \frac{S_{21}}{1 - \frac{S_{11}(z_r - 1)}{z_r + 1}} \tag{6}$$

Since the relative permittivity, relative permeability and absorptivity are obtained by $\varepsilon_r = \frac{n_r}{z_r}$, $\mu_r = n_r z_r$ and $A = 1 - |S_{11}|^2 - |S_{21}|^2$, high absorption means minimizing $|S_{11}|$ and $|S_{21}|$. Moreover, the impedance-matching condition $z_r = 1$ or $n_r = \sqrt{\varepsilon_r \mu_r} = \varepsilon_r = \mu_r$ for perfect absorption should be satisfied:

$$\begin{cases} \frac{\mu_r}{\varepsilon_r} = 1 \\ Im(\varepsilon_r)kd = Im(\mu_r)kd \gg 1 \end{cases} \tag{7}$$

The first equation in (7) is the impedance-matching condition, for minimizing the reflection from the surface of the metamaterial absorber. The second one means the dissipation inside should be large enough to attenuate the energy of incidence. Since the wave vector is proportional to the frequency, a larger thickness is required to achieve the same absorption at a lower frequency. Theoretically, the thickness $d$ of absorber should not be less than the limitation calculated by the reflection coefficient spectrum $R(\lambda)$ [28]:

$$\left| \int_0^\infty \ln|R(\lambda)| \, d\lambda \right| \leq 2\pi^2 d \tag{8}$$

To minimize the practical thickness as close as possible to the theoretical limitation, the absorber should possess flat ultra-broadband dispersion constitutive parameters. According to the Kramer-Kronig relation [8],

$$\begin{cases} Re(\chi(\omega)) = 1 + \frac{2}{\pi \int_0^{+\infty} \frac{d\omega' Im(\chi(\omega'))\omega'}{\omega'^2 - \omega^2}} \\ Im(\chi(\omega)) = -2\omega/\pi \int_0^{+\infty} d\omega' [Re(\chi(\omega')) - 1]/(\omega'^2 - \omega^2) \end{cases} \tag{9}$$

there are certain relations between the real part and the imaginary part of the constitutive parameters $\chi(\omega)$ (permittivity or permeability). The Kramer-Kronig relation shows that the real/imaginary part of constitutive parameters are not independent can be changed by image/real part. So, the ultra-broadband flat dispersions relation can be achieved by deliberately introducing multiple resonances with special dissipation [22]. For dispersive dielectric materials, since the real part of the permittivity $\varepsilon_r$ is inversely proportional to the squared incident wavelength $\lambda^2$ [29], it is easier to achieve multiple resonance modes within adjacent frequency band for modifying the dispersion. Compared to metallic structures, the dispersive dielectric materials have considerable potential to achieve ultra-broadband flat dispersion with simpler structures.

## II. Coupled Resonance Mechanism of Broadband Absorption

### a) Dielectric Resonance Mode of Single Structure

From the viewpoint of scattering theory, all scattering objects can be equivalent to effective electric or/and magnetic polarizability densities [30]. The electric and magnetic resonance modes are both supported in dielectric materials [31]. For non-dispersive dielectric structures, the supported low-coupled low-order and high-order resonant modes cannot achieve ultra-broadband absorption [32], however, the coupled adjacent resonances can extend the absorption frequency band [33]. Since the real part of the permittivity of water is inversely proportional to the squared incident wavelength, the coupled adjacent resonant modes can be excited in a certain frequency band simultaneously for ultra-broadband impedance-matching condition. In addition, the extremely high loss tangent in microwave frequency band indicates high absorption [34]. For single resonant structure, the magnetic resonances are stronger than electric resonances, so the excited multiple low-order and high-order vortex current can achieve ultra-broadband high absorption [35].

Single type of resonant structures, such as spherical [26], water-droplet [24], and rectangular [34-36] designs, have relatively narrow absorption frequency band for poor excited resonant modes and low coupling strength. After optimization, a variety of adjacent strongly coupled resonant modes can be excited to achieve ultra-broadband absorption [37-45]. As shown in Fig. 2(b), the single water-droplet structure can support coupled strong resonance peak A and C to realize ultra-broadband absorption. Figure. 2(c-d) show low-order and high-order vortex currents corresponding to different resonant loops of magnetic resonances. Since the length of the current loop is inversely proportional to resonant frequency, higher frequency means shorter vortex current loops [37]. Limited by the designed structure, the magnetic resonances of single water-droplet structure have low coupling strength, which leads to several isolated absorption peak. As shown in Fig. 2(e), the optimized water-droplet structure with bottom cone can support more resonance modes, therefore the high-coupled adjacent peaks can achieve ultra-broadband absorption [45].

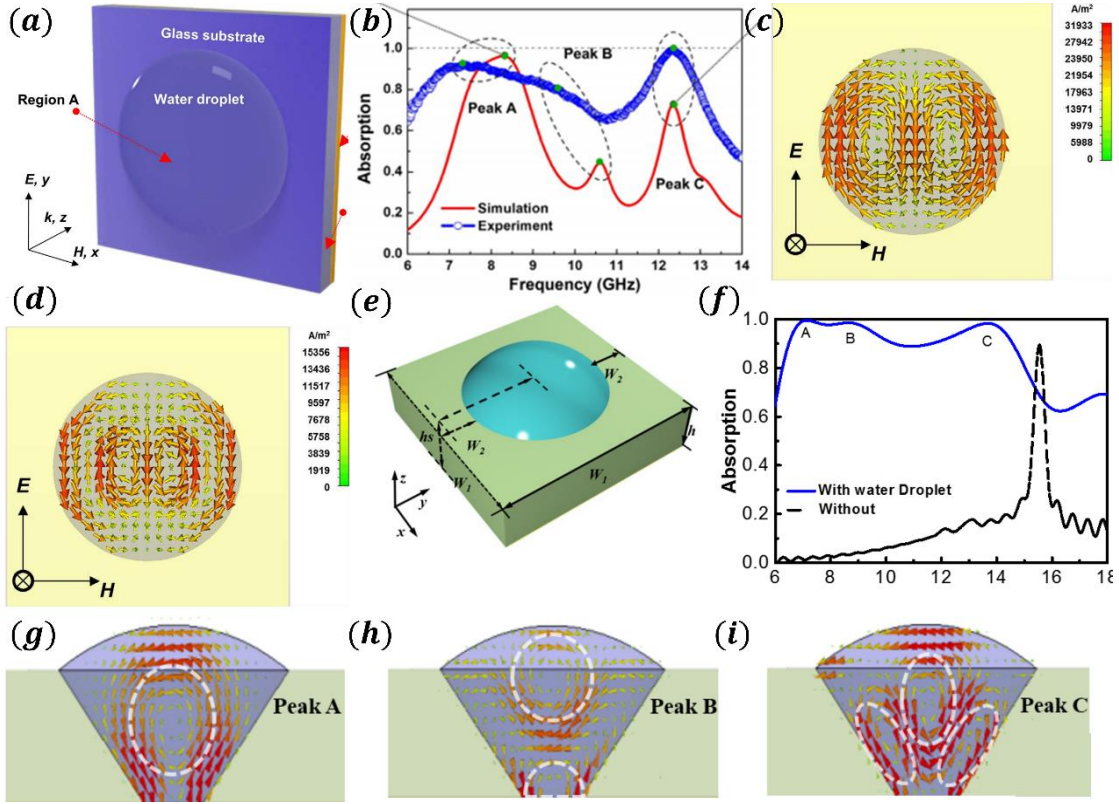

Fig. 2. The coupled dielectric resonance modes of WMMAs. (a) The 3D schematic diagram of droplet-shaped WMMAs. (b) The simulation and experiment results of absorption relations. The density distribution of induced current at resonance (c) peak A and (d) peak B. (e) The 3D schematic diagram and (f) absorption curves of optimized droplet-shaped WMMAs. (g-i) The vortex current distribution of optimized droplet-shaped WMMAs at resonance peak A, B and C. (a-d) Reproduced with permission. [24]Copyright 2015, Nature Springer Publishing. (e-i) Reproduced with permission. [45]Copyright 2018, OSA Publishing.

b) **Dielectric Resonance Mode of Horizontal and Vertical Integrated Structure**

According to the absorption superposition method [33], the horizontal-integrated resonant structures can achieve broader absorption frequency band by coupling the adjacent ones [45, 46]. As shown in Fig. 3(a), since the single water-tube structure support two adjacent resonant modes A and B, the total absorption frequency band is relative narrow. In contrast, as shown in Fig. 3(b), because the excited resonant peaks are highly connected to geometry parameters, the rectangular spiral shaped structure with extended absorption frequency band can be achieved by integrating water-tubes with different diameters [5].

Horizontal-integrated method not only leads to large period of the absorber unit cell, but is also difficult to excite the diffraction grating resonance mode for high absorption in higher frequency [47]. In order to achieve smaller absorber unit cell, structures with adjacent resonance frequency band can be superimposed vertically [48]. As shown in Fig. 3(f-g), the channel water-based absorber is composed of three-layered water-based structures in Model_IV. The small cross structure in Model_III improves the absorption in the middle frequency band (30-50 GHz), and the square ring structure with long resonant loop in Model_IV ameliorates the absorption in low frequency band (10-20 GHz). The vertical-coupled large cross structure, small cross structure, and square ring structure can achieve ultra-broadband high absorption [49]. As shown in Fig. 3(h), the water-based substrate coupled with tapered metal structure can also support multiple resonance modes to achieve ultra-broadband absorption. With the increased resonant frequency, the long resonant loop at the bottom of the structure changes to the shorter resonant loop in the top structure [50]. As illustrated in Fig. 3(i), due to the variable aspects ratio, the water-based tapered moth-eye structure can also support multiple resonance modes to achieve ultra-broadband absorption [39].

Most vertically superimposed water-based absorbers have a large thickness after forward design and local optimization [50], and utilize the tapered structure for ultra-broadband absorption [51]. According to Rozanov limit [28], both the larger degrees of freedom in optimization degree and more deliberately designed optimization algorithm are required to reduce the thickness close to the theoretical limit [23]. Under the same design complexity, the initial absorbing frequency of water-based structure with dispersion characteristics is lower than metallic structures [39, 50].

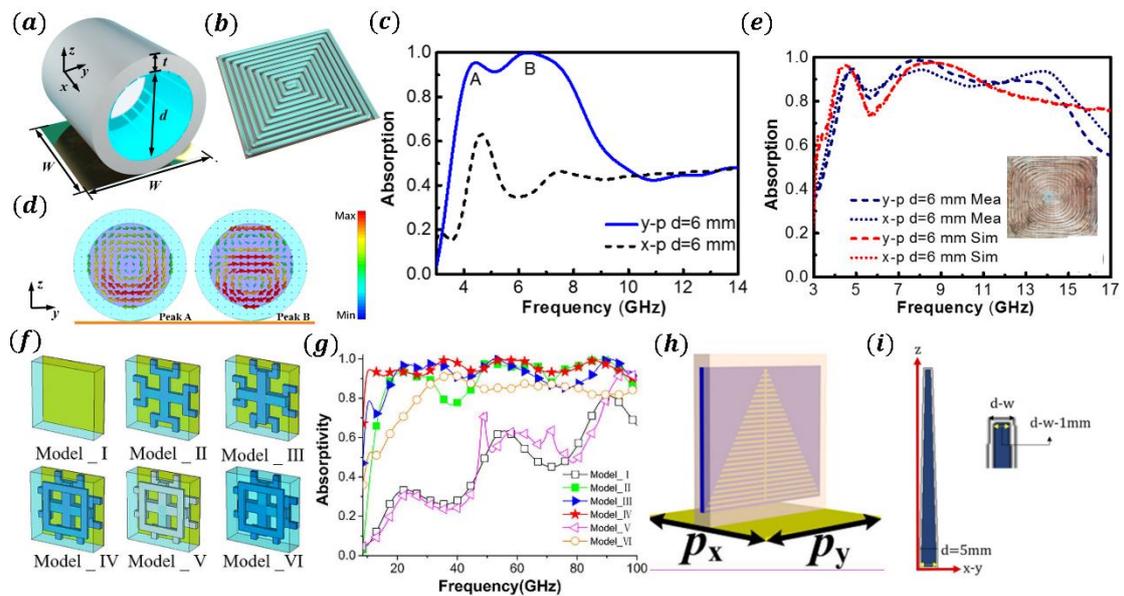

Fig. 3. The Horizontal and Vertical Integrated Structure water-based material absorbers. The 3D schematic diagram

of (a) tube-shaped water-based absorber and (b) horizontal integrated tube-shaped water-based absorber. (c) The absorption of tube-shaped water-based absorber for y-polarized and x-polarized normal incidence. (d) The vortex current loop density distributions for at resonance peak A and B. (e) The experimental and simulated absorptivity of horizontal integrated tube-shaped water-based absorber. (f) The vertical integrated design processing diagrams of channeled water-based absorber. (g) The absorptivity curves of Model (I-VI). (h) The schematic diagram of fish-bone shaped metal and layered water absorber. (i) The schematic diagram of glass caped moth-eye water-based absorber. (a-e) Reproduced with permission. [45]Copyright 2018, OSA Publishing. (f-g) Reproduced with permission. [49]Copyright 2021, OSA Publishing. (h) Reproduced with permission. [50]Copyright 2018, Nature Springer Publishing. (i) Reproduced with permission. [39]Copyright 2021, OSA Publishing.

### c) Diffraction Gratings and Surface Plasmonic Polarizations Modes

Through the horizontal or vertical broadband integration method, we can extend the absorption frequency band, and improve the low-frequency absorption. Due to the constitutive high loss tangent, the high-frequency absorption of WMMA with two-dimensional grating arrangement mainly depends on the excited two-dimensional diffraction grating mode [47]. The water-based absorbing structure with periodic length $p$ and refractive index $n$ can support zero-order diffraction grating mode and first-order diffraction grating mode (higher order modes are negligible), while the zero-order diffraction can excite the dielectric resonance mode. When incident wavelength $\lambda$ satisfies Eq. (10),

$$\frac{\lambda}{n} < p < \lambda \tag{10}$$

the first-order diffraction grating mode can be excited, the diffraction angle $\theta$ of which can be described by Eq. (11) as

$$\sin(\theta) = \frac{\lambda}{np}. \tag{11}$$

Due to the large value of loss tangent of water in high frequency, the high density at the top of the water-based absorbing structure can be achieved by first-order diffraction grating mode. When the absorber has metal substrate to prevent backscattering, the horizontal wave vector component of first-order diffraction grating mode can excite the confined surface plasmonic polarizations mode in dielectric-metal surface. The relation between horizontal component of the incident wave vector $k_x$ and SPP mode wave vector $k_{SPP}$ satisfies Eq. (12) [52],

$$k_{SPP} = k_x \pm m\frac{2\pi}{p}, m = 0, \pm 1, \pm 2 \cdots. \tag{12}$$

However, the intensity of the excited SPP resonant mode is closely related to the frequency. In the high frequency, the diffraction component of incidence is absorbed at the top of water-based absorbing structure, so the excited SPP resonant mode is weak. When the frequency decreases, the diffraction component excites strong SPP resonant mode in dielectric-metal confined surface.

As shown in Fig. 4(b-e), four high absorption resonance peaks are supported in the typical two-dimensional grating WMMA in Fig. 4(a). Since the metal substrate prevents transmission at low frequency, the high absorption at frequency $f_1$ and $f_2$ is mainly due to the dielectric resonance and metal substrate interaction. When the frequency increases, the high absorption at $f_3$ depends on the SPP resonant mode excited by the diffraction grating mode, while extremely high absorption in the top at $f_4$ is achieved by the dominant diffraction grating mode [53]. Considering the ground-free case, the first-order diffraction grating mode can only exist at higher frequency of 20 GHz. Moreover, the magnetic and electric component distributions in Fig. 4(i-j) demonstrate that the SPP resonant mode can be effectively excited by diffraction grating mode [54].

Only WMMA with periodic two-dimensional gap distribution can support the excited diffraction grating mode. Therefore, the WMMA with no obvious periodic gap distribution is difficult to excite the two-dimensional diffraction mode for high-frequency absorption [29, 38, 42, 43]. According to Rozanov limit, the correlation between the thickness limit and high-frequency absorption is weak, so it is easy to achieve a thinner absorber in high frequency band [28]. By introducing two-dimensional grating distribution structure in WMMAs, the high-frequency absorption is effectively improved by utilizing the diffraction grating mode and corresponding SPP resonant mode [37, 53, 54]. In addition, the high absorption of ground-free WMMA in high frequency can also be achieved by introducing the two-dimensional grating design [39, 40].

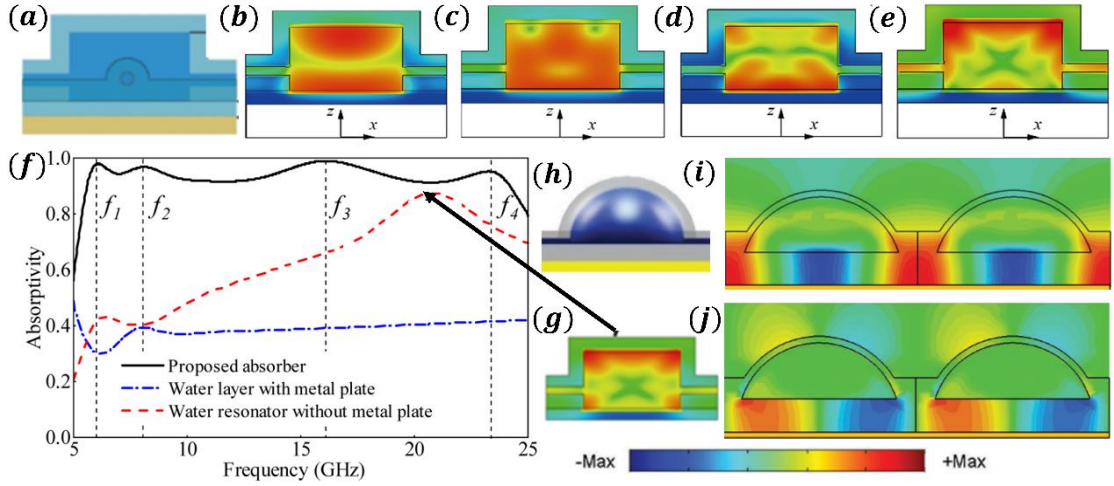

Fig. 4. The diffraction gratings and surface plasmonic polarization modes of WMMA. (a) The 3D schematic diagram of cylindrical WMMA. (b-e) The energy loss density distributions of cylindrical WMMA at resonance peaks $f_1$, $f_2$, $f_3$ and $f_4$. (f) The absorptivity comparison of proposed absorber, water layer with metal plate and water resonator without metal plate. (g) The energy loss distribution of ground-free water-based absorber with diffraction grating mode. (h) The 3D diagram of sphere-cap WMMA. (i) The magnetic component $H_x$ and (j) electric component $E_z$ of sphere-cap water-based absorber for diffraction grating mode. (a-g) Reproduced with permission. [53]Copyright 2018, OSA Publishing. (h-j) Reproduced with permission. [54]Copyright 2017, Wiley Publishing.

## III. Polarization- and Angle- Insensitivity

All of the WMMAs with multiple symmetry are characterized with both polarization insensitivity and high absorption rate to the incident wave [37-39, 41-43]. A typical example is the two-layer irregular WMMA with quadruple rotational symmetry, which can achieve perfect polarization-insensitive absorption as shown in Fig. 5(a).

For the oblique incidence condition, the excited resonance intensity of water-based dielectric structure is relatively sensitive to the horizontal component of magnetic field [5, 10, 44]. For larger incident angle, the resonance intensity decreases obviously under oblique TE incidence due to the decreased horizontal component, while the resonance intensity decreases slightly under oblique TM incidence due to the unchanged parallel magnetic field. As shown in Fig. 5(c-d), the WMMAs can still obtain high absorption under an incident angle as large as 60°, which can be applied in wide-angle absorption [49].

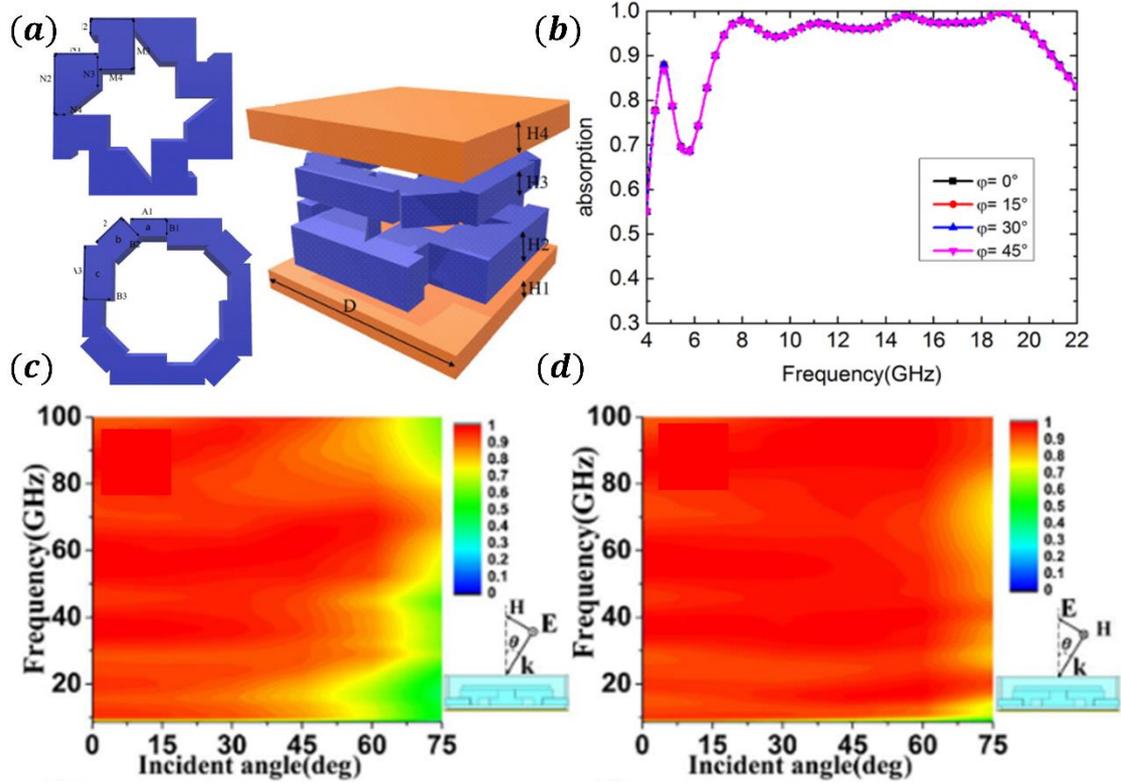

Fig. 5. The polarization- and angle- insensitive absorption of WMMA. (a) The 3D schematic diagram of irregular water-based absorber. (b) The simulation result of absorption with different polarization angles. The absorptivity of channeled WMMA with oblique (c) TE incident wave and (d) TM incident wave. (a-b) Reproduced with permission. [43]Copyright 2018, AIP Publishing. (c-d) Reproduced with permission. [49]Copyright 2021, OSA Publishing.

## 1. Tunable Absorption

### I. Thermal tunability

As shown in Eq. (13-15), the Debye model parameters of water, including optical permittivity $\varepsilon_\infty(T)$, static permittivity $\varepsilon_0(T)$, and relaxation time $\tau(T)$, can be described as a function of temperature [25]. Since polarity of liquid water decreases due to hydrogen bond fracture after heating, the real and imaginary parts of permittivity, as well as the loss tangent angle $tan\delta$ within 0-40 GHz, decrease with higher temperature as exhibited in Fig. 6(b). For dielectric resonance in WMMA, the lower $tan\delta$ means lower absorption intensity, while the variable real part of permittivity achieve the offset of resonance peak.

$$\varepsilon_0(T) = a_1 - b_1 T + c_1 T^2 - d_1 T^3 \qquad (13)$$

$$\varepsilon_\infty(T) = \varepsilon_0(T) - a_2 e^{-b_2 T} \qquad (14)$$

$$\tau(T) = c_2 e^{\frac{d_2}{T+T_0}} \tag{15}$$

The proposed absorber composed of metal-pattern and water-layer is illustrated in Fig. 6(c), where the water layer works as the dielectric lossy layer. Moreover, the permittivity $\varepsilon$ of water layer can be represented as the lumped element capacitance C in resonant circuit with parallel connection. Therefore, it is effective to adjust the amplitude and phase of resonance by the thermal control of water layer [55]. The variation proportion of imaginary part in Fig. 6(d) is larger than real part under increasing temperature, achieving large change of loss tangent. Therefore, the absorption intensity at the coupling of resonance peaks decreases significantly due to weaker resonance peaks, showing promising thermal tunable absorption applications [50].

Besides, the absorption with thermal insensitivity also has novel application. As shown in Fig. 6(b), the loss tangent of water is sensitive to temperature within 0-10 GHz, so it is difficult to realize thermal stable absorption design [36]. Most of the thermal stable WMMAs developed so far only works in higher frequency band (>10 GHz) [37, 40, 53]. Coupling with the thermal-insensitive SPP mode and diffraction grating mode, the thermal stable absorption of WMMA can be realized.

When there is a metal substrate, the WMMA in Fig. 6(e) can also support SPP mode. Compared with absorber composed of metal-pattern and water layer, the WMMA has high thermal stability due to the coupled strong diffraction grating and SPP mode [53]. In addition, the high thermal stable absorption can also be achieved by coupling the diffraction grating mode and dielectric mode in ground-free structure [40]. Further works on how to achieve thermal stable absorption at lower frequency are still challenging.

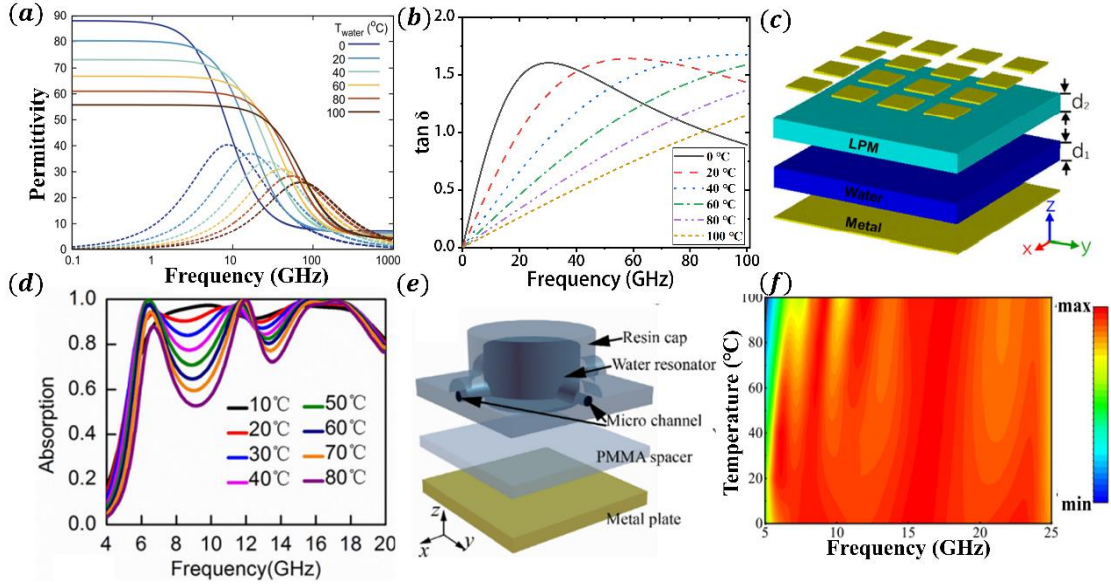

Fig. 6. The thermal tunability and stability of WMMAs. (a) The permittivity of water as a function of frequency for the temperature 0-100 °C. (The solid lines correspond to the real part, dashed lines to the imaginary part). (b) The loss tangent $tan\delta$ as a function of frequency for the temperature range of 0-100 °C. (c) The 3D schematic diagram of thermal tunable metamaterial absorber. (d) The absorption of thermal tunable metamaterial absorber for different temperatures varying from 10°C to 80°C. (e) The 3D schematic diagram of thermal stable metamaterial absorber. (f) The absorption relations of thermal stable metamaterial absorber for different temperatures varying from 0°C to 100°C. (a) Reproduced with permission. [26]Copyright 2015, Nature Springer Publishing. (b) Authors' unpublished figure. (c-d) Reproduced with permission. [55]Copyright 2016, AIP Publishing. (e-f) Reproduced with permission. [53]Copyright 2018, OSA Publishing.

## II. Ionic tunability

According to Debye relations Eq. (1-4), the loss tangent of water at low frequency is small, so the WMMA cannot achieve high absorption due to the low energy loss of resonance loop. The loss tangent of water at low frequency can be effectively improved by adding ions to form an ionic solution for higher low-frequency energy loss. Taking the common ion NaCl as an example, the modified Debye model of saline is [56]:

$$\varepsilon = \varepsilon_\infty + \frac{\varepsilon_s(T,N) - \varepsilon_\infty}{1 - i\omega\tau(T,N)} + i\frac{\sigma_{NaCl}(T,N)}{\omega\varepsilon_0^*} \qquad (16)$$

Compared with unmodified Debye model, $i\frac{\sigma_{NaCl}(T,N)}{\omega\varepsilon_0^*}$ is added to the imaginary part of permittivity. As shown in Fig. 7(b), saline has larger imaginary part of permittivity in the low frequency band than water, and thus the metamaterial absorber based on saline has higher low-frequency absorption under the same structure. As shown in Fig. 7(e), when the normalized salinity $S$ is 0.25, the absorption of circular truncated cone in Fig. 7(c) at low frequency is greatly improved. The Figure. 7(d) shows that the strong dielectric resonance loss at low frequency solves the common problem of low absorption below 3 GHz in WMMAs [57].

When the frequency increases, the permittivity of water has low correlation with salinity, so the saline-based metamaterial absorber can be regarded as pure water-based structure. Due to the existence of multiple resonance modes in higher frequency band, such as dielectric resonance mode, SPP mode and diffraction grating mode, the ions have less effect on absorption in higher frequency range.

As shown in Fig. 7(e), the tunable absorption can be achieved by adjusting the salinity in low frequency band. Inspired by thermal tunable WMMA, in order to reduce the influence of SPP mode and diffraction grating mode for high-frequency sensitive tunable absorption, the ionic-WMMA composed of water layer and metal structure are proposed. Further works of ionic tunable absorption are expected on the multi-manipulation with electric method such as programmable tunable control [58]. The programmable beam control absorption structure shown in Fig. 7 (f) can adjust the amplitude and phase of different spectrum reflection peaks corresponding to the on and off states of the diode [59]. The multiple tunable absorption applications need to be further studied.

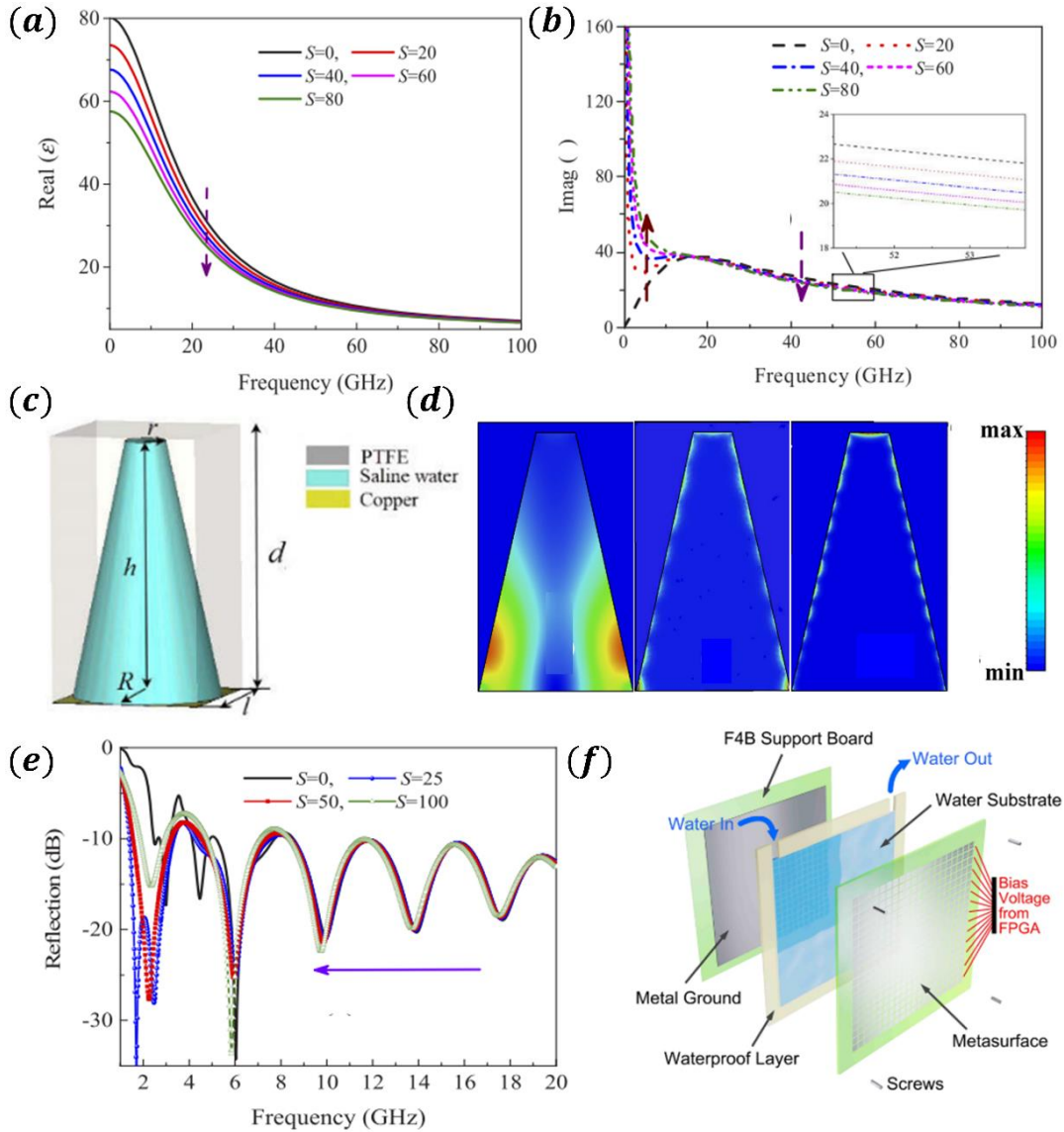

Fig. 7. The ionic tunable absorption of WMMAs. (a) The real part and (b) imaginary part of permittivity $\varepsilon$ as a function of frequency for different normalized salinity. (c) The 3D schematic diagram of saline tunable metamaterial absorber. (d) The loss density distributions of saline tunable metamaterial absorber from low to high frequencies. (1.73, 30.6 and 50.2 GHz) (e) The absorption of saline tunable metamaterial absorber for different salinity. (f) The schematic ionic tunable absorber with programmable design. (a-e) Reproduced with permission. [57]Copyright 2020, OSA Publishing. (f) Reproduced with permission. [59]Copyright 2019, AIP Publishing.

## III. Shape Tunability

Compared with absorbers of solid state, the WMMAs have unique tunability to control the absorption due to the liquid state. The developed shape-control design methods are divided into two categories: one is to use water pipes or microfluidic systems to control the water content in the structure [54, 60, 61], and the other is to reserve part of the space in WMMA to control the shape by gravity [26, 62].

As shown in Fig. 8(a), the shape of WMMA is tuned by controlling the internal water pressure in the PDMS deformable container with microfluidic system. When water pressure is high, the increased volume brings longer dielectric resonance loop with longer path and dielectric resonance peak with lower frequency. While the diffraction grating mode and SPP resonance mode keep unchanged, so the lower coupling between adjacent resonant peaks leads to lower absorption [54]. As shown in Fig. 8 (b), for the metal-water-based structure, the position of the resonance peak can be effectively adjusted by controlling the thickness of the water layer. The increased thickness of water layer leads to red shift and blue shift of low-frequency and high-frequency resonance peaks, respectively, due to the decreased thickness of the air layer above [60].

The second gravity-controlled tunable route depends on the angle between the structure and the gravity direction. However, when the gravity-tunable plane of the overall structure is perpendicular to the ground, the tunability will be lost. Moreover, the fine-tuning control cannot be achieved by gravity. As shown in Fig. 8(c), half of the cavity filled with water can achieve different shapes and different transmission properties by rotating the overall structure [62]. Figure. 8(d) shows a possible tunable process of absorbing switch. An hourglass-like water-based structure can reach two states of high and low absorption through the principle of connector [26].

By utilizing the external fluidic systems and gravity, the shape of water-based absorbing structure varies. However, fluidic control changes the water content, while gravity regulation does not. For the tunable method with unchanged water content, inspired by origami mechanical force control [63], the water-based material can be bound through an elastic container, and the mechanical deformation of the structure in Fig. 8 (e) can be realized by using external mechanical force to adjust the absorption.

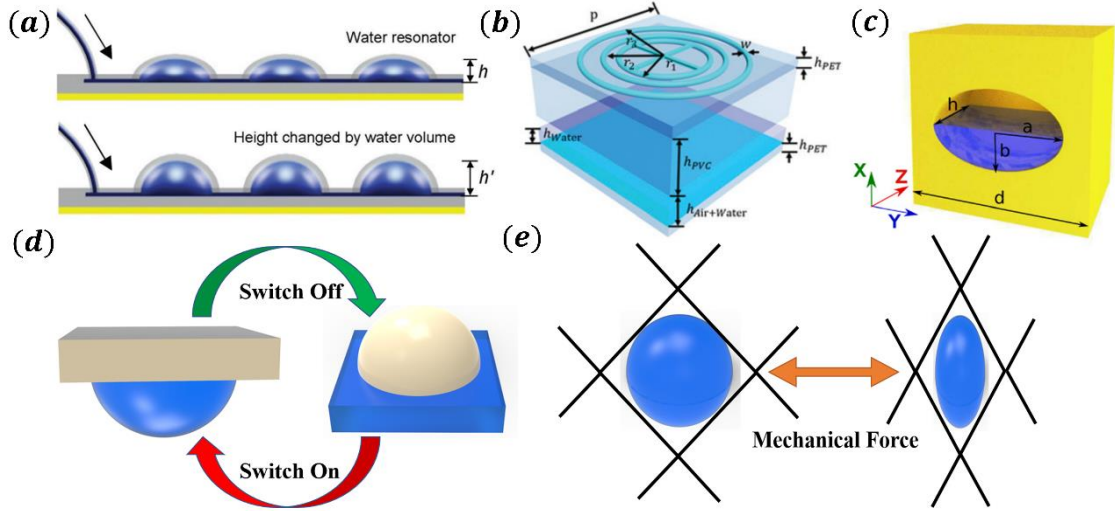

Fig. 8. The shape-tunable WMMAs. (a) The 3D schematic diagram of micro-channel shape-controlled water-based absorber. (b) The 3D schematic diagram of layer-tunable thick-controlled water-based absorber. (b) The 3D schematic diagram of gravity-controlled rotational tunable water-based absorber. (d) The schematic diagram of gravity-controlled switchable water-based absorber. (The blue and white structures are water and container, respectively.) (e) The foldable mechanical-controlled shaped-tunable WMMA. (a) Reproduced with permission. [54]Copyright 2017, Wiley Publishing. (b) Reproduced with permission. [60]Copyright 2021, RSC Publishing. (c) Reproduced with permission. [62]Copyright 2016, AIP Publishing. (d-e) Authors' unpublished figures.

## 2. Multi-functional Absorption

### I. Optical Transparent Function

Optical transparent absorbers with high ultra-broadband microwave absorption have application value in camouflage, energy shielding and photovoltaic system. The dispersion characteristics of water in near infrared and visible frequencies can be described by Sellmeier equation (extinction coefficient is ignored) [64]:

$$n = \sqrt{1 + \frac{0.75831\lambda^2}{\lambda^2 - 0.01007} + \frac{0.08495\lambda^2}{\lambda^2 - 8.91377}} \quad (17)$$

As shown in Fig. 9(a), the refractive index of water in the near-infrared and visible frequency bands is close to 1, so the water is optical transparent with high loss tangent in microwave frequency. Therefore, optical transparent broadband microwave water-based absorber can be realized by combining optical transparent materials with water [39, 60, 61, 65, 66].

The dielectric container of WMMA can be fabricated with PMMA, glass, PET and other optical transparent dielectric material [67, 68], while the metallic substrate and pattern layer can be replaced by printable optical transparent metal ITO [69]. As exhibited in Fig. 9(c), near and distant views of WMMA in Fig. 9(b) shows that the absorber have promising performance in optical transmission experimentally. As shown in Fig. 9(d), the comparison with PMMA-based metamaterial absorber proves that the WMMA is optical transparent with 60% optical transmissivity [60, 61]. In order to realize complex design with optical transparent property, we can choose transparent resin as the material of container for 3D printing.

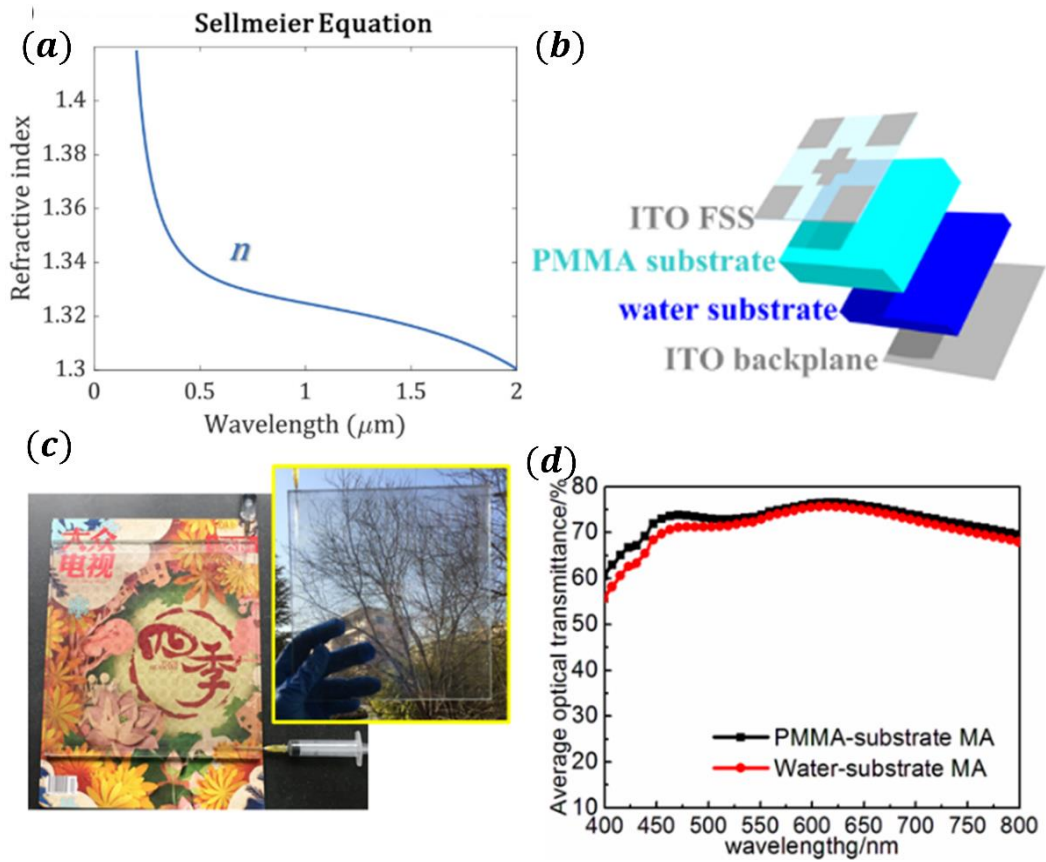

Fig. 9. The optical transparent function of WMMAs. (a) The refractive index $n$ calculated by Sellmeier equation in the wavelength range from 0.2 to 2 $\mu m$. (b) The 3D schematic diagram of typical optical transparent WMMA. (c) The experimental demonstration of near and distant views for optical transparent WMMA. (d) The comparison between PMMA-substrate metamaterial absorber and water-substrate metamaterial absorber. (a-d) Reproduced with permission. [61]Copyright 2018, OSA Publishing.

## II. Flexible Structure

Because PMMA, resin and other deformable dielectric materials are not suitable for flexible structure, the WMMAs in Fig. 10(a) can be fabricated by soft dielectric material and thin metal layer, such as PDMS and copper [54]. As shown in Fig. 10(c), the reflection spectrum of different directions of curved WMMA shows a high omnidirectional absorption. Intuitively, when a curved surface is illuminated under normal plane wave, the reflection wave scatters into different directions. As illustrated in Fig. 10(b), the bare metal shows relatively high omnidirectional reflection spectrum. This comparison indicates that the water-based structure contributes a lot to the omnidirectional absorption.

Due to limitation relation between liquid water and dielectric container, the geometry of water-based structure varies with the curved flexible base structure, so the connected water layer structure may have uneven overall distribution. Therefore, the base structure deformation should be considered in the design of flexible water-based structure, and only specific water-based absorber is suitable for flexible design [29].

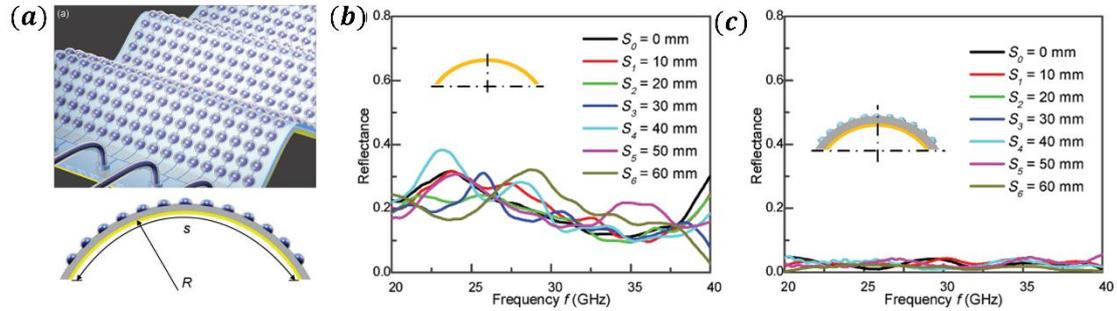

Fig. 10. The flexible WMMA. (a) The schematic diagram and of flexible WMMA. The reflectance spectrum at different reflective positions for (b) bare metal curved surface and (c) curved water-based absorber when R=200 mm. (a-c) Reproduced with permission. [54]Copyright 2017, Wiley Publishing.

### III. Tunable Infrared Radiation

The infrared radiation of material follows the relation $M = \varepsilon(T)\sigma T^4$, where $\varepsilon(T)$ is infrared emissivity, and $T$ is the temperature. The dynamically tunable radiation property of materials in infrared frequency has introduced a wide range of novel applications [70, 71]. The most convenient route for control radiation is changing the chemical composition or kinetic temperature. Since the water has large specific heat capacity, the corresponding infrared radiation of the WMMAs can be effectively tuned by the injected water [60, 61].

As shown in Fig. 11(b-d), the infrared radiation of the WMMA is lower than that of PMMA-based absorber, thus can be controlled by injected water [65]. However, the absorption in low frequency band will change simultaneously with the variable structure of water layer. Further works on WMMA with tunable infrared radiation and stable microwave absorption are expected.

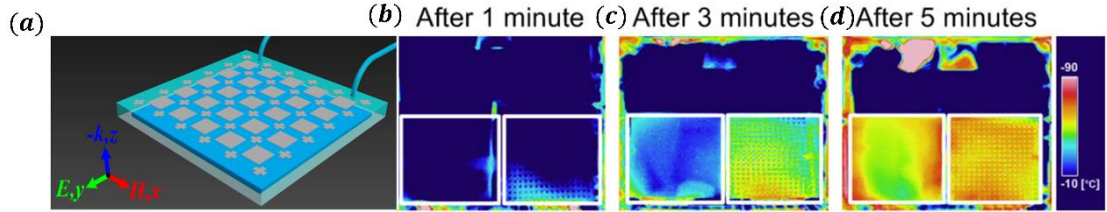

Fig. 11. The infrared-radiation tunable WMMA. (a) The schematic diagram of water-based absorber with control channels. (b-d) The IR radiation patterns of water-substrate absorber (left side) and PMMA-substrate absorber (right side) with initial heated temperature 45℃ in room temperature after 1, 3 and 5 min. Reproduced with permission. [61]Copyright 2018, OSA Publishing.

Table. 1 The comparison between selected ultra-broadband microwave water-based absorbers.

| Absorber Structure | Absorption band (GHz) | Thickness (mm) | RB[a] | RT[b] | Tunability/ Multifunction |
|---|---|---|---|---|---|
| [24]Water droplet | 8.3-12.07 or 9.2-16.5 | 4 | 57% | 0.111 | No |
| [44]Hollow layer | 12-29.6 | 5.8 | 85% | 0.232 | No |
| [42] H-shape | 7.9-21.7 | 5.8 | 93% | 0.153 | No |
| [72] Channel | 8.1-22.9 | 5.6 | 95% | 0.151 | No |
| [45]Tube | 5-15 | 7 | 100% | 0.117 | No |
| [40] Cuboid-layer | 7.74-23.56 | 12.8 | 101% | 0.330 | No |
| [43]Irregular layers | 6.8-21 | 10.4 | 102% | 0.236 | No |
| [41]Hollow-cylinder and layer | 6.5-21.4 | 10.6 | 106% | 0.230 | No |
| [53]Cylinder | 5.58-24.21 | 5.6 | 125% | 0.104 | No |
| [46]Swastika-shape | 9.3-49 | 3.6 | 136% | 0.112 | No |
| [37]Hollow-cylinder | 6.96-72.8 | 3.8 | 165% | 0.088 | No |
| [49]Micro channel | 9.6-98.9 | 3 | 165% | 0.096 | No |
| [55]copper pattern and water layer | 6.2-19 | 3.5 | 101% | 0.072 | Thermal tunability |
| [50]Fish-bone copper and water substrate | 2.6-16.8 | 15 | 146% | 0.130 | Thermal tunability |
| [54]Sphere-cap | 12-40 | 1.6 | 107% | 0.064 | Shape tunability |
| [60]ITO pattern and water layer | 5.8-16.2 | 4.15 | 94% | 0.080 | Shape tunability |
| [38] Step-structure | 5.5-27.5 | 5.8 | 133% | 0.106 | Flexible structure |
| [29]Optimized layer | 5.9-25.6 | 4 | 134% | 0.079 | Flexible structure |
| [65] Cylinder-layer | 6.4-30 | 4.5 | 130% | 0.096 | Optical transparent |
| [39]Moth-eye | 4-120 | 55 | 187% | 0.733 | Optical transparent |
| [61]ITO pattern and water layer | 6.4-23.7 | 3.675 | 115% | 0.078 | Shape and IR radiation tunability, Optical transparent |

| [57]Saline cone | 1.4-3.3 and 4.3-63 | 26 | 174% | 0.121 | Salinity tunability |

[a] The relative bandwidth is defined as RB = 100% *2( $f_{max}$ − $f_{min}$)/( $f_{max}$ + $f_{min}$), where $f_{max}$ and $f_{min}$ are the lowest and highest frequency of the absorption band, respectively.

[b] The RT is the thickness of the absorber versus wavelength at the lowest frequency of the absorption.

## Outlook

As a dispersive material for t microwave metamaterial absorbers, water-based materials show serval natural advantages, notably in terms of the high loss tangent in microwave frequency band, and the flat dispersion relation in ultra-broadband absorption like metallic structures [22]. Although most WMMAs show a promising absorbing ability, the inverse design method lacks of theoretical instruction [73]. Further works are expected at goal-oriented inverse design with data driven route for WMMAs [74], so the WMMAs with thickness approaching theoretical limit can be achieved. Although water have low loss tangent in low frequency (1-5 GHz) for absorption, this can be improved by either adding ionic in the water for larger conductivity or replacing the water with ionic liquid [75]. Due to the large constitutive loss tangent in high frequency band, the WMMA can easily achieve large absorption in high frequency band. Towards the challenging theoretical relative bandwidth limit (200%), we can achieve this goal by utilizing the optimized water-based materials. Further explorations are expected in the tunable and multi-functional WMMAs. The thermal tunable, electric tunable and shape tunable WMMA can be fabricated by introducing phase change material [76] and electric tunable material [17]. The mildly changed material parameter of water combined with the abruptly changed one of phase change material provides versatile tunable functions. Based on tunable design, the WMMAs with multiple functions, such as wavefront control [77], polarization conversion combined with absorption [78, 79] and so on, can be realized. Moreover, the easy-to-get and cheap water and complex container fabricated by 3D printing technology show a promising potential in practical application. At last, further broadband functions and other applications can be explored by the constitutive property of water-based materials [80-86].

## Acknowledgement

This work is supported by the National Natural Science Foundation of China (No. 51872154). The authors gratefully acknowledge discussions with Dr. M. Wang, Dr. K. J. Yang, and Dr. P. D.

Yang.

## Conflict of interest

The authors declare that there is no conflict of interest during preparation of this manuscript.

## Reference

[1] Y. Qing, W. Zhou, S. Huang, Z. Huang, F. Luo and D. Zhu, Microwave absorbing ceramic coatings with multi-walled carbon nanotubes and ceramic powder by polymer pyrolysis route, Compos. Sci. Technol. 89 (2013) 10-14. http://doi.org/10.1016/j.compscitech.2013.09.007

[2] H. Nan, Y. Qing, H. Gao, H. Jia, F. Luo and W. Zhou, Synchronously oriented Fe microfiber & flake carbonyl iron/epoxy composites with improved microwave absorption and lightweight feature, Compos. Sci. Technol. 184 (2019) http://doi.org/10.1016/j.compscitech.2019.107882

[3] Y. Qing, W. Zhou, F. Luo and D. Zhu, Titanium carbide (MXene) nanosheets as promising microwave absorbers, Ceram. Int. 42 (2016) 16412-16416. http://doi.org/10.1016/j.ceramint.2016.07.150

[4] Y. Qing, W. Zhou, F. Luo and D. Zhu, Epoxy-silicone filled with multi-walled carbon nanotubes and carbonyl iron particles as a microwave absorber, Carbon. 48 (2010) 4074-4080. http://doi.org/10.1016/j.carbon.2010.07.014

[5] J. Wen, Q. Ren, R. Peng and Q. Zhao, An all-dielectric metasurface absorber based on surface wave conversion effect, Appl. Phys. Lett. 119 (2021) http://doi.org/10.1063/5.0057529

[6] D. R. Smith, D. C. Vier, S. C. Nemat-Nasser, and S. Schultz, Composite Medium with Simultaneously Negative Permeability and Permittivity, Phys. Rev. Lett. 84 (2000) http://doi.org/10.1103/PhysRevLett.84.4184

[7] J. B. Pendry, W. J. Stewart and I. Youngs, Extremely Low Frequency Plasmons in Metallic Mesostructures, Phys. Rev. Lett. 76 (1996) http://doi.org/10.1103/PhysRevLett.76.4773

[8] J. B. Pendry, D. J. Robbins, W. J. Stewart, Magnetism from Conductors and Enhanced Nonlinear Phenomena, IEEE Trans. Microwave Theory Tech. 47 (1999) 2075 - 2084. http://doi.org/10.1109/22.798002

[9] N. I. Landy, S. Sajuyigbe, J. J. Mock, D. R. Smith and W. J. Padilla, Perfect metamaterial absorber, Phys. Rev. Lett. 100 (2008) 4. http://doi.org/10.1103/PhysRevLett.100.207402

[10] X. M. Liu, K. Bi, B. Li, Q. Zhao and J. Zhou, Metamaterial perfect absorber based on artificial dielectric "atoms", Opt. Express. 24 (2016) 20454-20460. http://doi.org/10.1364/oe.24.020454

[11] X. M. Liu, C. W. Lan, K. Bi, B. Li, Q. Zhao and J. Zhou, Dual band metamaterial perfect absorber based on Mie resonances, Appl. Phys. Lett. 109 (2016) 5. http://doi.org/10.1063/1.4960802

[12] Q. Y. Wen, H. W. Zhang, Y. S. Xie, Q. H. Yang and Y. L. Liu, Dual band terahertz metamaterial absorber: Design, fabrication, and characterization, Appl. Phys. Lett. 95 (2009) http://doi.org/10.1063/1.3276072

[13] J. W. Park, P. V. Tuong, J. Y. Rhee, K. W. Kim, W. H. Jang, E. H. Choi, L. Y. Chen and Y. Lee, Multi-band metamaterial absorber based on the arrangement of donut-type resonators, Opt. Express. 21 (2013) 9691-9702. http://doi.org/10.1364/oe.21.009691

[14] Z. Y. Li, B. Li, Q. Zhao and J. Zhou, A metasurface absorber based on the slow-wave effect,

AIP Advances. 10 (2020) 5. http://doi.org/10.1063/1.5143408

[15]  Y. Zhi Cheng, Y. Wang, Y. Nie, R. Zhou Gong, X. Xiong and X. Wang, Design, fabrication and measurement of a broadband polarization-insensitive metamaterial absorber based on lumped elements, J. Appl. Phys. 111 (2012) http://doi.org/10.1063/1.3684553

[16]  J. Zhang, X. Z. Wei, I. D. Rukhlenko, H. T. Chen and W. R. Zhu, Electrically Tunable Metasurface with Independent Frequency and Amplitude Modulations, ACS Photonics. 7 (2020) 265-271. http://doi.org/10.1021/acsphotonics.9b01532

[17]  Q. Zhao, L. Kang, B. Du, B. Li, J. Zhou, H. Tang, X. Liang and B. Z. Zhang, Electrically tunable negative permeability metamaterials based on nematic liquid crystals, Appl. Phys. Lett. 90 (2007) http://doi.org/10.1063/1.2430485

[18]  Z. X. Su, Q. Zhao, K. Song, X. P. Zhao and J. B. Yin, Electrically tunable metasurface based on Mie-type dielectric resonators, Sci. Rep. 7 (2017) 1-7. http://doi.org/10.1038/srep43026

[19]  H. Pan and H. F. Zhang, Thermally tunable polarization-insensitive ultra-broadband terahertz metamaterial absorber based on the coupled toroidal dipole modes, Opt. Express. 29 (2021) 18081-18094. http://doi.org/10.1364/oe.427554

[20]  Q. Zhao, B. Du, L. Kang, H. Zhao, Q. Xie, B. Li, X. Zhang, J. Zhou, L. Li and Y. Meng, Tunable negative permeability in an isotropic dielectric composite, Appl. Phys. Lett. 92 (2008) http://doi.org/10.1063/1.2841811

[21]  P. Yu, L. V. Besteiro, Y. Huang, J. Wu, L. Fu, H. H. Tan, C. Jagadish, G. P. Wiederrecht, A. O. Govorov and Z. Wang, Broadband Metamaterial Absorbers, Adv. Opt. Mater. 7 (2018) http://doi.org/10.1002/adom.201800995

[22]  D. Ye, Z. Wang, K. Xu, H. Li, J. Huangfu, Z. Wang and L. Ran, Ultrawideband dispersion control of a metamaterial surface for perfectly-matched-layer-like absorption, Phys. Rev. Lett. 111 (2013) 187402. http://doi.org/10.1103/PhysRevLett.111.187402

[23]  S. Qu, Y. Hou and P. Sheng, Conceptual-based design of an ultrabroadband microwave metamaterial absorber, Proc Natl Acad Sci U S A. 118 (2021) http://doi.org/10.1073/pnas.2110490118

[24]  Y. J. Yoo, S. Ju, S. Y. Park, Y. Ju Kim, J. Bong, T. Lim, K. W. Kim, J. Y. Rhee and Y. Lee, Metamaterial Absorber for Electromagnetic Waves in Periodic Water Droplets, Sci Rep. 5 (2015) 14018. http://doi.org/10.1038/srep14018

[25]  W. J. Ellison, Permittivity of Pure Water, at Standard Atmospheric Pressure, over the Frequency Range 0–25THz and the Temperature Range 0–100°C, J. Phys. Chem. Ref. Data. 36 (2007) 1-18. http://doi.org/10.1063/1.2360986

[26]  A. Andryieuski, S. M. Kuznetsova, S. V. Zhukovsky, Y. S. Kivshar and A. V. Lavrinenko, Water: Promising Opportunities For Tunable All-dielectric Electromagnetic Metamaterials, Sci. Rep. 5 (2015) 9. http://doi.org/10.1038/srep13535

[27]  X. D. Chen, T. M. Grzegorczyk, B. I. Wu, J. Pacheco and J. A. Kong, Robust method to retrieve the constitutive effective parameters of metamaterials, Phys. Rev. E. 70 (2004) http://doi.org/10.1103/PhysRevE.70.016608

[28]  K. N. Rozanov, Ultimate Thickness to Bandwidth Ratio of Radar Absorbers, IEEE Trans. Antennas Propag. 48 (2000) 1230-1234. http://doi.org/10.1109/8.884491

[29]  Z. Wu, X. Chen, Z. Zhang, L. Heng, S. Wang and Y. Zou, Design and optimization of a flexible water-based microwave absorbing metamaterial, Appl. Phys. Express. 12 (2019) http://doi.org/10.7567/1882-0786/ab0f66

[30] M. Albooyeh, D. Morits and S. A. Tretyakov, Effective electric and magnetic properties of metasurfaces in transition from crystalline to amorphous state, Phys. Rev. B. 85 (2012) http://doi.org/10.1103/PhysRevB.85.205110

[31] Q. Zhao, L. Kang, B. Du, H. Zhao, Q. Xie, X. Huang, B. Li, J. Zhou and L. Li, Experimental demonstration of isotropic negative permeability in a three-dimensional dielectric composite, Phys. Rev. Lett. 101 (2008) 027402. http://doi.org/10.1103/PhysRevLett.101.027402

[32] X. M. Liu, Q. Zhao, C. W. Lan and J. Zhou, Isotropic Mie resonance-based metamaterial perfect absorber, Appl. Phys. Lett. 103 (2013) 3. http://doi.org/10.1063/1.4813914

[33] N. L. Mou, S. L. Sun, H. X. Dong, S. H. Dong, Q. He, L. Zhou and L. Zhang, Hybridization-induced broadband terahertz wave absorption with graphene metasurfaces, Opt. Express. 26 (2018) 11728-11736. http://doi.org/10.1364/oe.26.011728

[34] D. J. Gogoi and N. S. Bhattacharyya, Embedded dielectric water "atom" array for broadband microwave absorber based on Mie resonance, J. Appl. Phys. 122 (2017) http://doi.org/10.1063/1.4995519

[35] D. J. Gogoi and N. S. Bhattacharyya, Metasurface absorber based on water meta "molecule" for X-band microwave absorption, J. Appl. Phys. 124 (2018) http://doi.org/10.1063/1.5041450

[36] X. Sun, Q. Fu, Y. Fan, H. Wu, K. Qiu, R. Yang, W. Cai, N. Zhang and F. Zhang, Thermally controllable Mie resonances in a water-based metamaterial, Sci Rep. 9 (2019) 5417. http://doi.org/10.1038/s41598-019-41681-5

[37] X. Zhang, F. Yan, X. Du, W. Wang and M. Zhang, Broadband water-based metamaterial absorber with wide angle and thermal stability, AIP Advances. 10 (2020) http://doi.org/10.1063/5.0006166

[38] Y. Zhou, Z. Shen, J. Wu, Y. Zhang, S. Huang and H. Yang, Design of ultra-wideband and near-unity absorption water-based metamaterial absorber, Appl. Phys. B. 126 (2020) http://doi.org/10.1007/s00340-020-7401-y

[39] H. Kwon, G. D'Aguanno and A. Alu, Optically transparent microwave absorber based on water-based moth-eye structures, Opt. Express. 29 (2021) 9190-9198. http://doi.org/10.1364/OE.418220

[40] J. Xie, S. Quader, F. Xiao, C. He, X. Liang, J. Geng, R. Jin, W. Zhu and I. D. Rukhlenko, Truly All-Dielectric Ultrabroadband Metamaterial Absorber: Water-Based and Ground-Free, IEEE Antennas Wirel. Propag. Lett. 18 (2019) 536-540. http://doi.org/10.1109/lawp.2019.2896166

[41] S. Li, Z. Shen, H. Yang, Y. Liu, Y. Yang and L. Hua, Ultra-wideband Transmissive Water-Based Metamaterial Absorber with Wide Angle Incidence and Polarization Insensitivity, Plasmonics, (2021) http://doi.org/10.1007/s11468-021-01389-7

[42] Y. Zhou, Z. Shen, X. Huang, J. Wu, Y. Li, S. Huang and H. Yang, Ultra-wideband water-based metamaterial absorber with temperature insensitivity, Phys. Lett. A. 383 (2019) 2739-2743. http://doi.org/10.1016/j.physleta.2019.05.050

[43] Z. Shen, X. Huang, H. Yang, T. Xiang, C. Wang, Z. Yu and J. Wu, An ultra-wideband, polarization insensitive, and wide incident angle absorber based on an irregular metamaterial structure with layers of water, J. Appl. Phys. 123 (2018) http://doi.org/10.1063/1.5024319

[44] J. Xie, W. Zhu, I. D. Rukhlenko, F. Xiao, C. He, J. Geng, X. Liang, R. Jin and M. Premaratne, Water metamaterial for ultra-broadband and wide-angle absorption, Opt. Express. 26 (2018) 5052-5059. http://doi.org/10.1364/OE.26.005052

[45] J. Zhao, S. Wei, C. Wang, K. Chen, B. Zhu, T. Jiang and Y. Feng, Broadband microwave

absorption utilizing water-based metamaterial structures, Opt. Express. 26 (2018) 8522-8531. http://doi.org/10.1364/OE.26.008522

[46] X. Zhang, D. Zhang, Y. Fu, S. Li, Y. Wei, K. Chen, X. Wang and S. Zhuang, 3-D Printed Swastika-Shaped Ultrabroadband Water-Based Microwave Absorber, IEEE Antennas Wirel. Propag. Lett. 19 (2020) 821-825. http://doi.org/10.1109/lawp.2020.2981405

[47] M. Pu, C. Hu, C. Huang, Z. Zhao, Y. Wang, and X. Luo, Engineering heavily doped silicon for broadband absorber in the terahertz regime, Opt. Express. 20 (2012) http://doi.org/10.1364/OE.20.025513

[48] F. Ding, Y. Cui, X. Ge, Y. Jin and S. He, Ultra-broadband microwave metamaterial absorber, Appl. Phys. Lett. 100 (2012) http://doi.org/10.1063/1.3692178

[49] Y. Chen, K. Chen, D. Zhang, S. Li, Y. Xu, X. Wang and S. Zhuang, Ultrabroadband microwave absorber based on 3D water microchannels, Photonics. Res. 9 (2021) http://doi.org/10.1364/prj.422686

[50] Y. Shen, J. Zhang, Y. Pang, L. Zheng, J. Wang, H. Ma and S. Qu, Thermally Tunable Ultra-wideband Metamaterial Absorbers based on Three-dimensional Water-substrate construction, Sci Rep. 8 (2018) 4423. http://doi.org/10.1038/s41598-018-22163-6

[51] Z. Shen, S. Li, Y. Xu, W. Yin, L. Zhang and X. Chen, Three-Dimensional Printed Ultrabroadband Terahertz Metamaterial Absorbers, Phys. Rev. Appl. 16 (2021) http://doi.org/10.1103/PhysRevApplied.16.014066

[52] Y. Qu, Q. Li, H. Gong, K. Du, S. Bai, D. Zhao, H. Ye and M. Qiu, Spatially and Spectrally Resolved Narrowband Optical Absorber Based on 2D Grating Nanostructures on Metallic Films, Adv. Opt. Mater. 4 (2016) 480-486. http://doi.org/10.1002/adom.201500651

[53] J. Ren and J. Y. Yin, Cylindrical-water-resonator-based ultra-broadband microwave absorber, Opt. Mater. Express. 8 (2018) http://doi.org/10.1364/ome.8.002060

[54] Q. Song, W. Zhang, P. C. Wu, W. Zhu, Z. X. Shen, P. H. J. Chong, Q. X. Liang, Z. C. Yang, Y. L. Hao, H. Cai, H. F. Zhou, Y. Gu, G.-Q. Lo, D. P. Tsai, T. Bourouina, Y. Leprince-Wang and A.-Q. Liu, Water-Resonator-Based Metasurface: An Ultrabroadband and Near-Unity Absorption, Adv. Opt. Mater. 5 (2017) http://doi.org/10.1002/adom.201601103

[55] Y. Pang, J. Wang, Q. Cheng, S. Xia, X. Y. Zhou, Z. Xu, T. J. Cui and S. Qu, Thermally tunable water-substrate broadband metamaterial absorbers, Appl. Phys. Lett. 110 (2017) http://doi.org/10.1063/1.4978205

[56] A. Stogryn, Equations for Calculating the Dielectric Constant of Saline Water (Correspondence), IEEE Trans. Microwave Theory Tech. 19 (1971) 733-736.

[57] H. Xiong and F. Yang, Ultra-broadband and tunable saline water-based absorber in microwave regime, Opt. Express. 28 (2020) 5306-5316. http://doi.org/10.1364/OE.382719

[58] T. J. Cui, M. Q. Qi, X. Wan, J. Zhao and Q. Cheng, Coding metamaterials, digital metamaterials and programmable metamaterials, Light Sci. Appl. 3 (2014) http://doi.org/10.1038/lsa.2014.99

[59] L. Chen, H. L. Ma, Y. Ruan and H. Y. Cui, Dual-manipulation on wave-front based on reconfigurable water-based metasurface integrated with PIN diodes, J. Appl. Phys. 125 (2019) http://doi.org/10.1063/1.5078660

[60] Y. Zhang, H. Dong, N. Mou, H. Li, X. Yao and L. Zhang, Tunable and transparent broadband metamaterial absorber with water-based substrate for optical window applications, Nanoscale. 13 (2021) 7831-7837. http://doi.org/10.1039/d0nr08640a

[61] Y. Shen, J. Zhang, Y. Pang, J. Wang, H. Ma and S. Qu, Transparent broadband metamaterial

[61] absorber enhanced by water-substrate incorporation, Opt. Express. 26 (2018) 15665-15674. http://doi.org/10.1364/OE.26.015665

[62] M. Odit, P. Kapitanova, A. Andryieuski, P. Belov and A. V. Lavrinenko, Experimental demonstration of water based tunable metasurface, Appl. Phys. Lett. 109 (2016) http://doi.org/10.1063/1.4955272

[63] J. T. B. Overvelde, T. A. de Jong, Y. Shevchenko, S. A. Becerra, G. M. Whitesides, J. C. Weaver, C. Hoberman and K. Bertoldi, A three-dimensional actuated origami-inspired transformable metamaterial with multiple degrees of freedom, Nat. Commun. 7 (2016) http://doi.org/10.1038/ncomms10929

[64] Q. George M. Hale, Optical Constants of Water in the 200-nm to 200-Mm Wavelength Region, Appl. Opt. 12 (1973) 555-563. http://doi.org/10.1364/AO.12.000555

[65] Y. Pang, Y. Shen, Y. Li, J. Wang, Z. Xu and S. Qu, Water-based metamaterial absorbers for optical transparency and broadband microwave absorption, J. Appl. Phys. 123 (2018) http://doi.org/10.1063/1.5023778

[66] Q. Wang, K. Bi and S. Lim, All-Dielectric Transparent Metamaterial Absorber With Encapsulated Water, IEEE Access. 8 (2020) 175998-176004. http://doi.org/10.1109/access.2020.3026358

[67] S. Huang, Q. Fan, C. Xu, B. Wang, J. Wang, B. Yang, C. Tian and Z. Meng, Multiple working mechanism metasurface with high optical transparency, low infrared emissivity and microwave reflective reduction, Infrared. Phys. Techn. 111 (2020) http://doi.org/10.1016/j.infrared.2020.103524

[68] C. Xu, B. Wang, M. Yan, Y. Pang, W. Wang, Y. Meng, J. Wang and S. Qu, An optical-transparent metamaterial for high-efficiency microwave absorption and low infrared emission, J. Phys. D: Appl. Phys. 53 (2020) http://doi.org/10.1088/1361-6463/ab651a

[69] S. Zhong, L. Wu, T. Liu, J. Huang, W. Jiang and Y. Ma, Transparent transmission-selective radar-infrared bi-stealth structure, Opt. Express. 26 (2018) 16466-16476. http://doi.org/10.1364/OE.26.016466

[70] T. Inoue, M. De Zoysa, T. Asano and S. Noda, Realization of dynamic thermal emission control, Nature Materials. 13 (2014) 928-931. http://doi.org/10.1038/nmat4043

[71] X. Y. Liu and W. J. Padilla, Reconfigurable room temperature metamaterial infrared emitter, Optica. 4 (2017) 430-433. http://doi.org/10.1364/optica.4.000430

[72] X. J. Huang, H. L. Yang, Z. Y. Shen, J. Chen, H. I. Lin and Z. T. Yu, Water-injected all-dielectric ultra-wideband and prominent oblique incidence metamaterial absorber in microwave regime, J. Phys. D. 50 (2017) 7. http://doi.org/10.1088/1361-6463/aa81af

[73] H. Rajabalipanah, A. Abdolali and M. Mohammadi, Experimental and analytical investigations on a wide-angle, polarization-insensitive, and broadband water-based metamaterial absorber, J. Phys. D: Appl. Phys. 54 (2021) http://doi.org/10.1088/1361-6463/abe3af

[74] W. Ma, F. Cheng and Y. M. Liu, Deep-Learning-Enabled On-Demand Design of Chiral Metamaterials, ACS Nano. 12 (2018) 6326-6334. http://doi.org/10.1021/acsnano.8b03569

[75] E. Yang, F. Yang, J. Pei, X. Zhang, S. Liu and Y. Deng, All-dielectric ultra-broadband metamaterial absorber based on imidazole ionic liquids, J. Phys. D: Appl. Phys. 52 (2019) http://doi.org/10.1088/1361-6463/ab2d9a

[76] L. Lei, F. Lou, K. Y. Tao, H. X. Huang, X. Cheng and P. Xu, Tunable and scalable broadband metamaterial absorber involving VO2-based phase transition, Photonics. Res. 7 (2019) 734-741.

http://doi.org/10.1364/prj.7.000734

[77] Y. Shen, J. Zhang, S. Sui, Y. Jia, Y. Pang, J. Wang, H. Ma and S. Qu, Transparent absorption-diffusion-integrated water-based all-dielectric metasurface for broadband backward scattering reduction, J. Phys. D: Appl. Phys. 51 (2018) http://doi.org/10.1088/1361-6463/aae2fe

[78] H. He, X. Shang, L. Xu, J. Zhao, W. Cai, J. Wang, C. Zhao and L. Wang, Thermally switchable bifunctional plasmonic metasurface for perfect absorption and polarization conversion based on VO2, Opt. Express. 28 (2020) 4563-4570. http://doi.org/10.1364/oe.385900

[79] D. Yan, M. Meng, J. Li, J. Li and X. Li, Vanadium dioxide-assisted broadband absorption and linear-to-circular polarization conversion based on a single metasurface design for the terahertz wave, Opt. Express. 28 (2020) 29843-29854. http://doi.org/10.1364/oe.404829

[80] M. Askari, D. A. Hutchins, R. L. Watson, L. Astolfi, L. Nie, S. Freear, P. J. Thomas, S. Laureti, M. Ricci, M. Clark and A. T. Clare, An ultrasonic metallic Fabry–Pérot metamaterial for use in water, Addit Manuf. 35 (2020) http://doi.org/10.1016/j.addma.2020.101309

[81] B. X. Khuyen, V. T. H. Hanh, B. S. Tung, V. D. Lam, Y. J. Kim, Y. Lee, H.-T. Tu and L. Y. Chen, Narrow/Broad-Band Absorption Based on Water-Hybrid Metamaterial, Crystals. 10 (2020) http://doi.org/10.3390/cryst10050415

[82] R. E. Jacobsen, A. V. Lavrinenko and S. Arslanagic, A Water-Based Huygens Dielectric Resonator Antenna, IEEE Open Journal of Antennas and Propagation. 1 (2020) 493-499. http://doi.org/10.1109/ojap.2020.3021802

[83] A. Keshavarz and Z. Vafapour, Water-Based Terahertz Metamaterial for Skin Cancer Detection Application, IEEE Sens. J. 19 (2019) 1519-1524. http://doi.org/10.1109/jsen.2018.2882363

[84] M. L. P. Bailey, A. T. Pierce, A. J. Simon, D. T. Edwards, G. J. Ramian, N. I. Agladze and M. S. Sherwin, Narrow-Band Water-Based Absorber With High Return Loss for Terahertz Spectroscopy, IEEE Trans Terahertz Sci Technol. 5 (2015) 961-966. http://doi.org/10.1109/tthz.2015.2477609

[85] Z. Shen, H. Yang, X. Huang and Z. Yu, Design of negative refractive index metamaterial with water droplets using 3D-printing, Journal of Optics. 19 (2017) http://doi.org/10.1088/2040-8986/aa8a4c

[86] W. Zhu, I. D. Rukhlenko, F. Xiao, C. He, J. Geng, X. Liang, M. Premaratne and R. Jin, Multiband coherent perfect absorption in a water-based metasurface, Opt. Express. 25 (2017) 15737-15745. http://doi.org/10.1364/OE.25.015737